\documentclass[aps,nofootinbib,onecolumn,citeautoscript]{revtex4-2}
\usepackage{amsmath,amssymb,amsfonts}
\usepackage{enumerate}
\usepackage{xcolor}
\usepackage[colorlinks=true,citecolor=violet,linkcolor=blue,linkbordercolor=cyan]{hyperref}
\usepackage{setspace}
\usepackage{enumerate}
\usepackage{graphicx}
\usepackage{float}
\usepackage{subcaption}

\renewcommand{\&}{\textup{\symbol{`\&}}}

\begin{document}
\title{Effect of Quantum Information Scrambling on Bound Entangled States}
\author{ Suprabhat Sinha$^\ast$\footnote[0]
{$^\ast$suprabhatsinha64@gmail.com}\\\vspace{0.3cm} \textit{School of Computer Science, Engineering and Applications \\D Y Patil International University, Akurdi, Pune-411044, India}}

\begin{abstract}
Spreading information in physical systems is a common phenomenon. However, when the information is quantum in nature, tracking, describing, and quantifying the information is a challenging task. Quantum information scrambling defines the quantum information propagating chaotically over the physical system. This article describes the effect of quantum information scrambling on bound entangled states. A bound entangled state is a particular type of entangled state that carries noisy entanglement. The distillation of this type of entangled state is very difficult. In recent times, the usefulness of these states has been depicted in different applications. The outcome of this study exhibits that quantum information scrambling develops entanglement in the separable portion of the bound entangled states. Although quantum information scrambling reduces entanglement, the study pointed out that quantum information scrambling plays a significant role in activating the bound entangled states by introducing a certain amount of approximately stable entanglement.
\end{abstract}

\maketitle

\section{Introduction}\label{sec:1}
Quantum Information (QI) scrambling addresses the quantum manifestation of the chaotic nature of classical information dynamics of systems. When a system interacts with another system, local information preserved in the initial system diffuses over the total system chaotically. It is very challenging to accumulate the information perfectly. If the information is quantum, it is impossible to pile up the quantum information using local measurements. The loss of quantum information due to local measurement, strictly speaking the amount of quantum information incapable of recovering using local measurement, is defined as QI scrambling. Explaining the black hole information paradox by showing that black holes rapidly process the quantum information and exhibit the fastest QI scrambling, Hayden \textit{et al.}\cite{hn} attracts the scientific community towards QI scrambling. After this, several researches are conducted applying QI scrambling in miscellaneous domains like condensed matter physics, high energy physics, information theory, quantum thermodynamics, and so on \cite{qs1,qs2,qs3,qs4,kk1}. Several diverse approaches have already been proposed to quantify this chaotically scrambled information in physically interacting systems like Loschmidt Echo, entropy production, out-of-time-ordered correlator (OTOC) \cite{le, ep,otoc1,otoc2} etc. Among these different varieties of quantifiers, OTOC has acquired the most attraction in recent years. 

On the other hand, quantum entanglement, a fundamental property of quantum particles, is one of the founding members of the branch of quantum computation and quantum information theory. From the beginning, entanglement shows its gravity in this branch and proves itself as a monumental asset. As the branch moves forward, the scientific community starts several studies on entangled quantum states from different points of the compass. Some research concludes that entangled quantum states can be split into two types. One type of entangled quantum states are distillable, and pure entanglement can be extracted easily. These types of entangled quantum states are termed as free entangled states. Another type of entangled quantum states are hard to distill and extract pure entanglement. These types of entangled quantum states are defined as bound entangled states \cite{be1,be2}. A diverse number of bound entangled states have already been proposed by different researchers. But for the requirement of maximum pure entanglement, free entangled states are preferred for perfect execution in most of the applications of this branch and bound entangled states are staved off. Some recent studies found that bound entangled states can be used in quantum information theory with some free entanglement\cite{qk,cri, tp, cc}. After that, a variety of research works have been conducted on dynamical analysis, distillation, and activation of bound entangled states\cite{gq,bz,kk2,ss1,ss2}. These research works involve a variety of methods for detecting and measuring entanglements. For free entangled states, several methods are available to quantify the free entanglement, such as concurrence, negativity, three $\pi$ measurement \cite{con, neg,3pi} etc. On the contrary, characterization and detection of the bound entanglement is still an open problem. Although some criteria have already been developed to detect the bound entanglement, such as separability criterion, realignment criterion, computable cross-norm or realignment (CCNR) criterion \cite{sc1,sc2,rc,ccnr} etc.

In the current article, the effect of QI scrambling on bound entangled states is cultivated. Although QI scrambling has already been studied in different qubit and qutrit systems and discussed the offshoot of QI scrambling on the respective considered systems. To the best of my knowledge, studying QI scrambling in different bound entangled states is lacking in the literature. This study is conducted on four $3\times3$ dimensional bipartite bound entangled quantum states provided by Bennett \textit{et al.}, Jurkowski \textit{et al.}, and Horodecki \textit{et al.} \cite{bs, js,hs1,hs2}. During the study, OTOC is applied to find the effect of QI scrambling on the bound entangled quantum states, negativity is employed for quantifying and measuring the free entanglement of the states, and the CCNR criterion is selected to detect the bound entanglement of the state.

This article is planned as follows. In section 2, OTOC and its role in QI scrambling, negativity, and CCNR criterion has been discussed. Section 3 dealt with the brief details of four chosen bound entangled states. In the different subsections of section 4, the effect of QI scrambling on bound entangled quantum states has been studied. The last section contains the conclusion of the study of this article.

\section{OTOC, Negativity and CCNR criterion}\label{sec:2}
In the current section, the role of OTOC in QI scrambling, negativity, and CCNR criterion has been discussed. OTOC is a commonly used quantifier for QI scrambling in the present day. It studies QI scrambling by measuring the degree of irreversibility of the system by applying the mismatch between the forward and backward evolution of the system. OTOC was first introduced by Larkin \textit{et al.} \cite{otoc1} as a quasiclassical method in superconductivity theory, and Hashimoto \textit{et al.} \cite{otoc2} introduced it in the field of quantum mechanics. The mathematical expression of OTOC can be written as,
\begin{equation}
f(t)=\langle \left[O_{2}(t),O_{1}\right]^{\dagger}\left[O_{2}(t),O_{1}\right] \rangle.\label{otoc}
\end{equation}
Where $O_{1}$ and $O_{2}(t)$ are the local operators which are  Hermitian as well as unitary. At initial time $(t=0)$, $O_{1}$ and $O_{2}(0)$ commute with each other (i.e. $\left[O_{2}(0),O_{1}\right]=0$). As the time moves forward, the operator $O_{1}$ remains unchanged, but the operator $O_{2}(t)$ evolves with time. Due to this evolution, the commutation relation between two operators generally crushes because of QI Scrambling. According to the Heisenberg Picture of quantum mechanics the operator $O_{2}(t)$ can be written as, $O_{2}(t)=U(t)^{\dagger}O_{2}(0)U(t)$. Where, $U(t)=e^{-\frac{iHt}{\hbar}}$ is the unitary time evolution operator under the Hamiltonian $H$. To calculate the QI scrambling of a quantum system with density matrix $\rho$, OTOC can be written as,
\begin{equation}
S(t)=Tr\left[\left(\left[O_{2}(t),O_{1}\right]^{\dagger}\left[O_{2}(t),O_{1}\right]\right)\rho\right].\label{scr}
\end{equation}
The above equation can be simplified as,
\begin{equation}
S(t)=2\left[1- Re\left(M \right)\right]\label{qis}.
\end{equation}
Where,
\begin{equation}
M=Tr\left[\rho(t)\right] \qquad \text{and}\qquad\rho(t)=O_{2}(t)  O_{1} O_{2}(t)  O_{1} \rho.\label{M}
\end{equation}

In the current study, $O_{1}$ is considered as a swap operator which swaps between qutrit $\vert 0 \rangle$, $\vert 2 \rangle$ and $O_{1}=O_{2}(0)$. It is also taken into account that $O_{2}(0)$ evolves with time under the Dzyaloshinskii-Moriya (DM) Hamiltonian in the Z-direction, which is developed due to the DM  interaction \cite{dm1,dm2,dm3} between the qutrits of the considered state. The mathematical expression of DM interaction Hamiltonian is written as,
\begin{equation}
H_{z}=D (\sigma_A^x \otimes \sigma_B^y - \sigma_A^y \otimes \sigma_B^x). \label{dm} 
\end{equation}
Where $D$ is the interaction strength along the Z-direction with the range $0 \leq D \leq 1$ and $\sigma_A^x$, $\sigma_A^y$ and $\sigma_B^x$, $\sigma_B^y$ are the spin matrices of qutrit $A$ and qutrit $B$ respectively. To simplify the calculations and discussion, $\hbar$ is assumed as 1 (i.e. $\hbar = 1$) throughout the present study. At $D=0$ there is no interaction between the qutrits of the considered state, so the operators $O_{1}$ and $O_{2}(t)$ commutes, and no QI scrambling takes place in the system. The matrix form of $O_{1}, \sigma^x$, $\sigma^y$ can be expressed as,
\begin{equation*}
O_{1}=\begin{bmatrix}
 0 & 0 & 1 \\
 0 & 1 & 0 \\
 1 & 0 & 0 \\
\end{bmatrix}
\qquad \sigma^x =
\begin{bmatrix}
 0 & \frac{1}{\sqrt{2}} & 0 \\
 \frac{1}{\sqrt{2}} & 0 & \frac{1}{\sqrt{2}} \\
 0 & \frac{1}{\sqrt{2}} & 0 \\
\end{bmatrix}
\qquad \sigma^y = 
\begin{bmatrix}
 0 & -\frac{i}{\sqrt{2}} & 0 \\
 \frac{i}{\sqrt{2}} & 0 & -\frac{i}{\sqrt{2}} \\
 0 & \frac{i}{\sqrt{2}} & 0 \\
\end{bmatrix}.
\end{equation*}

In the current work, negativity and CCNR criterion are used to detect and quantify the entanglement of the considered bound entangled state. To quantify the free entanglement of the state negativity has been employed, while the CCNR criterion has been applied to detect the bound entanglement of the state. CCNR is a simple and strong criterion for the separability of a density matrix. This criterion can detect a wide range of bound entangled states and performs with better efficacy. The negativity $(N)$ and CCNR criterion are defined as below,
\begin{equation}
N=\frac{(\left \|\rho_{AB}^{T}\right \|-1)}{2} \label{N}
\end{equation} 
and
\begin{equation}
CCNR=\left\|(\rho_{AB}-\rho_{A}\otimes \rho_{B})^R\right\|-\sqrt{(1-Tr \rho_{A}^{2}) (1-Tr \rho_{B}^{2})}. \label{CCNR}
\end{equation}
Where $\|...\|$, $(...)^T$, and $(...)^R$ represent the trace norm, partial transpose, and realignment matrix respectively. Further, $\rho_{AB}$ is the density matrix of the bound entangled state and $\rho_{A}$, $\rho_{B}$ are the reduced density matrices of qutrit A and qutrit B respectively and expressed as,
\begin{equation*}
\rho_{A}=Tr_{B}(\rho_{AB}) \quad \text{and}\quad   \rho_{B}=Tr_{A}(\rho_{AB}).
\end{equation*}
For a system, $N>0$ or $CCNR>0$ implies that the state is entangled, $N=0$ and $CCNR>0$ implicates that the state is bound entangled, and $N>0$ corresponds to the free entangled state.

\section{Bound entangled states} \label{sec:3}
In this section, the bound entangled states, which are studied in the current article, are cultivated. Bound entangled states are different types of entangled states that carry noisy entanglement, and it is hard to distill these types of entangled states. The usefulness of bound entangled states has been depicted in different applications. Many authors have already proposed different bound entangled states. Among them, four of the states have been chosen for this current study. The first considered bound entangled state is suggested by Bennett \textit{et al.} \cite{bs}. The state is a $3\times3$ dimensional bipartite bound entangled state dealing with two qutrits $A$ and $B$. The density matrix of the considered bound entangled state can be written in the form,

\begin{figure}[b]
\centering
	\begin{subfigure}{0.32\linewidth}
	\includegraphics[scale=0.52]{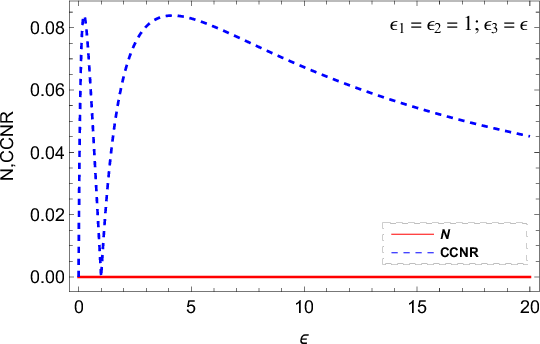} 
	\caption{For $\epsilon_{1}=\epsilon_{2}=1;\epsilon_{3}=\epsilon$}\label{f11}
	\end{subfigure}
	\begin{subfigure}{0.32\linewidth}
	\includegraphics[scale=0.52]{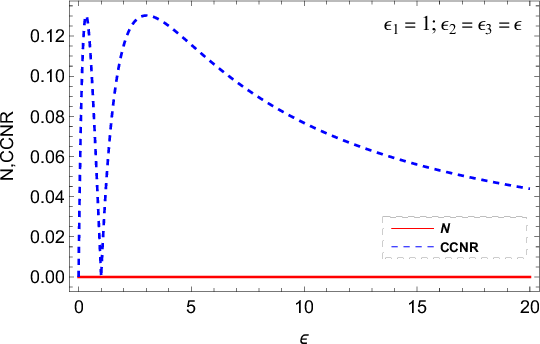} 
	\caption{For $\epsilon_{1}=1;\epsilon_{2}=\epsilon_{3}=\epsilon$}\label{f12}
	\end{subfigure}
	\begin{subfigure}{0.32\linewidth}
	\includegraphics[scale=0.52]{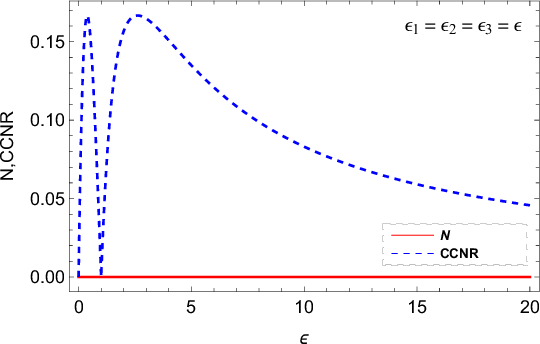}
	\caption{For $\epsilon_{1}=\epsilon_{2}=\epsilon_{3}=\epsilon$} \label{f13}
	\end{subfigure}
\caption{Plot of Negativity(N) and CCNR vs. $\epsilon$}\label{f1}
\end{figure}

\begin{equation}
\rho_{BS}=\frac{1}{4} \left[ (I \otimes I)-\sum_{i=0}^4\vert \psi_i\rangle \langle\psi_i \vert \right].
\label{bs}
\end{equation} 
Where, $I$ is the $3\times3$ dimensional identity matrix,
\begin{eqnarray*}
\vert\psi_0\rangle=\frac{1}{\sqrt{2}} \vert 0\rangle (\vert 0\rangle-\vert1\rangle),\quad
\vert\psi_1\rangle=\frac{1}{\sqrt{2}} (\vert0\rangle-\vert1\rangle)\vert2\rangle, 
\end{eqnarray*}
\begin{eqnarray*}
\vert\psi_2\rangle=\frac{1}{\sqrt{2}}\vert2\rangle (\vert1\rangle-\vert2\rangle), \quad
\vert\psi_3 \rangle=\frac{1}{\sqrt{2}}(\vert1\rangle-\vert2\rangle)\vert0\rangle, 
\end{eqnarray*}
\begin{eqnarray*}
\text{and} \quad
\vert\psi_4\rangle=\frac{1}{3} (\vert0\rangle+\vert1\rangle+\vert2\rangle)(\vert0\rangle+\vert1\rangle+\vert2\rangle).
\end{eqnarray*}

The second bound entangled state is proposed by Jurkowski \textit{et al.} \cite{js}. This is a parameterized, $3\times3$ dimensional bipartite bound entangled state constructed with qutrits $A$ and $B$. The chosen bound entangled state depends on the three parameters $\epsilon_{1}, \epsilon_{2}\text{ and }\epsilon_{3}$ and comes with different parameter conditions. When the parameters $\epsilon_{1}=\epsilon_{2}=\epsilon_{3}=1$ the state behaves like a separable state. The density matrix of the state can be written as,
\begin{equation}
\rho_{JS}=\frac{1}{N}
\begin{bmatrix}
 1 & 0 & 0 & 0 & 1 & 0 & 0 & 0 & 1 \\
 0 & \epsilon _{1} & 0 & 0 & 0 & 0 & 0 & 0 & 0 \\
 0 & 0 & \frac{1}{\epsilon _{3}} & 0 & 0 & 0 & 0 & 0 & 0 \\
 0 & 0 & 0 & \frac{1}{\epsilon_{1}} & 0 & 0 & 0 & 0 & 0 \\
 1 & 0 & 0 & 0 & 1 & 0 & 0 & 0 & 1 \\
 0 & 0 & 0 & 0 & 0 & \epsilon_{2} & 0 & 0 & 0 \\
 0 & 0 & 0 & 0 & 0 & 0 & \epsilon_{3} & 0 & 0 \\
 0 & 0 & 0 & 0 & 0 & 0 & 0 & \frac{1}{\epsilon_{2}} & 0 \\
 1 & 0 & 0 & 0 & 1 & 0 & 0 & 0 & 1 \\
\end{bmatrix}. \label{Js}
\end{equation}
Where, $N=\left(\epsilon _{1}+\frac{1}{\epsilon _{3}}+\frac{1}{\epsilon_{1}}+\epsilon_{2}+\epsilon_{3}+\frac{1}{\epsilon_{2}}+3 \right)$.\\
Figure \ref{f1} has shown the entanglement behavior of the Jurkowski \textit{et al.} bound entangled state for the different conditions of the parameters $\epsilon_{1}, \epsilon_{2}\text{ and }\epsilon_{3}.$ 

\begin{figure}[b]
\centering
	\begin{subfigure}{0.49\linewidth}
	\includegraphics[scale=0.75]{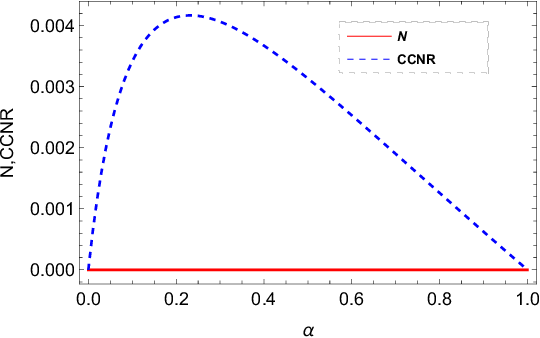} 
	\caption{For State 1}\label{f21}
	\end{subfigure}
	\begin{subfigure}{0.49\linewidth}
	\includegraphics[scale=0.75]{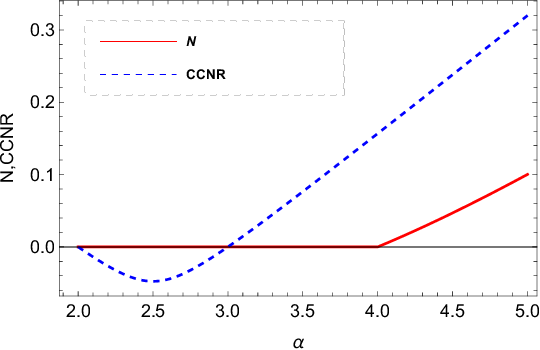}
	\caption{For State 2} \label{f22}
	\end{subfigure}
\caption{Plot of Negativity(N) and CCNR vs. $\alpha$}\label{f2}
\end{figure}

The third and fourth bound entangled states are investigated by Horodecki \textit{et al.} \cite{hs1,hs2}. Both of these states are also $3\times3$ dimensional bipartite bound entangled states formed by qutrits $A$ and $B$ with a parameter $\alpha$. The density matrix of one of the states investigated by Horodecki \textit{et al.} \textbf{[State 1]} is written as,
\begin{equation}
\rho_{HS_{1}}= 
\begin{bmatrix}
 \frac{\alpha }{8 \alpha +1} & 0 & 0 & 0 & \frac{\alpha }{8 \alpha +1} & 0
   & 0 & 0 & \frac{\alpha }{8 \alpha +1} \\
 0 & \frac{\alpha }{8 \alpha +1} & 0 & 0 & 0 & 0 & 0 & 0 & 0 \\
 0 & 0 & \frac{\alpha }{8 \alpha +1} & 0 & 0 & 0 & 0 & 0 & 0 \\
 0 & 0 & 0 & \frac{\alpha }{8 \alpha +1} & 0 & 0 & 0 & 0 & 0 \\
 \frac{\alpha }{8 \alpha +1} & 0 & 0 & 0 & \frac{\alpha }{8 \alpha +1} & 0
   & 0 & 0 & \frac{\alpha }{8 \alpha +1} \\
 0 & 0 & 0 & 0 & 0 & \frac{\alpha }{8 \alpha +1} & 0 & 0 & 0 \\
 0 & 0 & 0 & 0 & 0 & 0 & \frac{\alpha +1}{2 (8 \alpha +1)} & 0 &
   \frac{\sqrt{1-\alpha ^2}}{2 (8 \alpha +1)} \\
 0 & 0 & 0 & 0 & 0 & 0 & 0 & \frac{\alpha }{8 \alpha +1} & 0 \\
 \frac{\alpha }{8 \alpha +1} & 0 & 0 & 0 & \frac{\alpha }{8 \alpha +1} & 0
   & \frac{\sqrt{1-\alpha ^2}}{2 (8 \alpha +1)} & 0 & \frac{\alpha +1}{2
   (8 \alpha +1)} \\
\end{bmatrix}. \label{hs1}
\end{equation}
The range of the parameter $\alpha$ is $0 \leq \alpha \leq 1$ for the above-mentioned state (State 1).

\begin{figure}[b]
\centering
\includegraphics[scale=1]{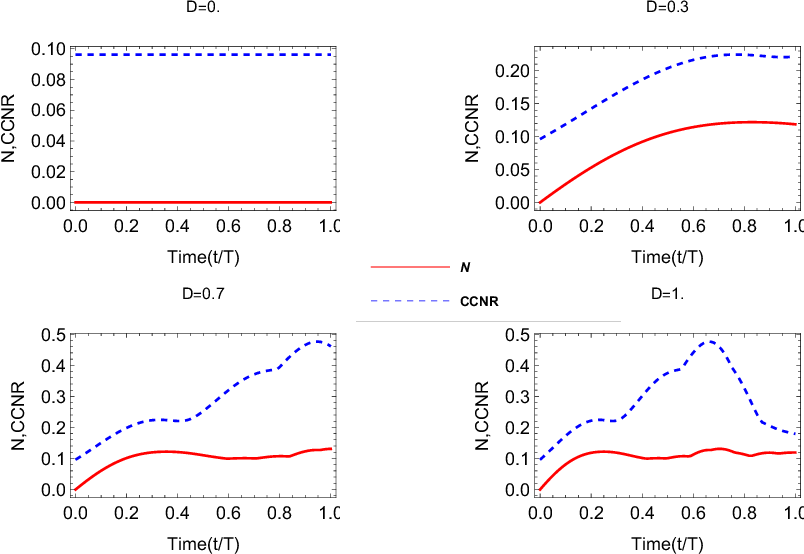} 
\caption{Plot of Negativity(N) and CCNR vs.Time$(t/T)$ for different values of $D$}\label{f3}
\end{figure}

The density matrix of the another state \textbf{[State 2]} can be written in the form as, 
\begin{equation}
\rho_{HS_{2}}= \frac{2}{7}\Delta+\frac{\alpha}{7}\delta^{+}+\frac{5-\alpha}{7}\delta^{-}.\label{hs2}
\end{equation}
Where,
\begin{eqnarray*}
\Delta&=&\vert \psi \rangle \langle \psi \vert \;; \quad \vert \psi \rangle =\frac{1}{\sqrt{3}}\left(\vert 00 \rangle +\vert 11 \rangle  +\vert 22 \rangle \right) ,\\
\delta^{+}&=& \frac{1}{3}\left(\vert 01 \rangle \langle 01 \vert+\vert 12 \rangle \langle 12 \vert+\vert 20 \rangle \langle 20 \vert \right),\\
\delta^{-}&=&  \frac{1}{3}\left(\vert 10 \rangle \langle 10 \vert+\vert 21 \rangle \langle 21 \vert+\vert 02 \rangle \langle 02 \vert \right).
\end{eqnarray*}
The discussed state (State 2) follows the parameter $\alpha$'s limit as $2 \leq \alpha \leq 5$ with the following conditions,
\begin{equation*}
\rho_{HS_{2}}=\left\lbrace \begin{array}{ll}
\text{Separable state for }& 2 \leq \alpha \leq 3 ,\\
\text{Bound entangled state for }& 3 < \alpha \leq 4 ,\\
\text{Free entangled state for }& 4 < \alpha \leq 5 ,\\
\end{array}\right ..
\end{equation*}
Figure \ref{f2} has depicted the entanglement behavior of both Horodecki \textit{et al.} bound entangled states with the parameter $\alpha$.

\begin{figure}[b]
\centering
\includegraphics[scale=1]{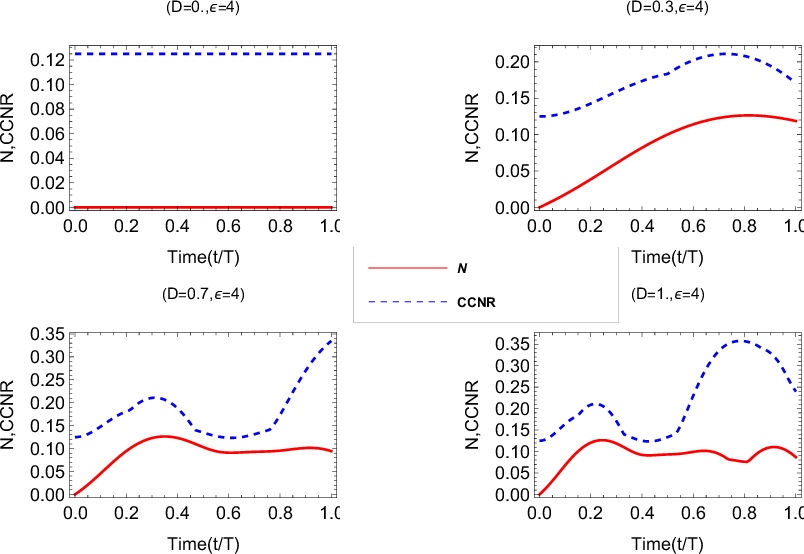} 
\caption{Plot of Negativity(N) and CCNR vs. Time$(t/T)$ for $\epsilon =4$ and different values of $D$}\label{f4}
\end{figure}

\section{Effect of QI scrambling on bound entangled states}
\label{sec:4}
In this section, the effect of QI scrambling on the selected bound entangled states has been discussed. During this study, the chosen bound entangled states have gone through the forward-backward evolution under the considered swap operators $O_{1}$ and $O_{2}(t)$. After passing through the evolution process, the density matrix of the evolved bound entangled state $(\rho(t))$ can be calculated using the equation \ref{M}. If the trace value of this evolved bound entangled state (i.e. the value of $M$ in equation \ref{M}) is $1$, then no QI is scrambling in the system. Since this study is focused on the effect of QI scrambling on bound entangled states, the density matrix $\rho(t)$ of the evolved bound entangled state has been cultivated in this article. Negativity and CCNR criteria are used in the present discussion for quantification and detection of entanglement as mentioned before. The present study is conducted for four different bound entangled states provided by three different authors and discusses the results in the three sequential cases, which are described in the consecutive subsections below.

\begin{figure*}
\centering
\includegraphics[scale=1]{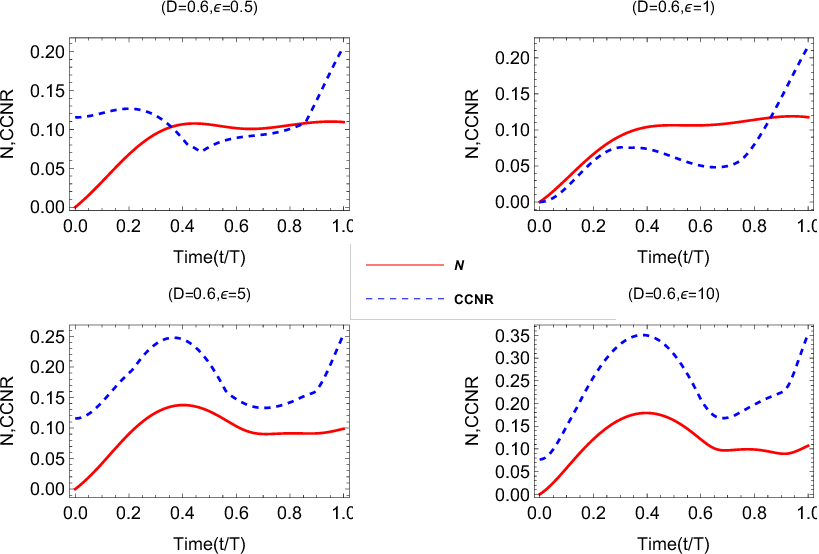} 
\caption{Plot of Negativity(N) and CCNR vs. Time$(t/T)$ for $D=0.6$ and different values of $\epsilon$}\label{f5}
\end{figure*}

\subsection*{Case 1: Effect on Bennett \textit{et al.} state}  
\label{sec:5}
Bennett \textit{et al.} provided a non-parameterized bound entangled state, which is discussed in equation \ref{bs}, is considered and dealt with in this current case. The state has been passed through the evolution process to examine the effect of QI scrambling on the state and shown the outcomes in figure \ref{f3} for the different grades of the interaction strength $D$. In the figure, the negativity (N) of the system is pointed out by the solid red line, and the CCNR criterion of the system is illustrated by the dashed blue line, which will be followed throughout this article.

The figure shows that for the interaction strength $D=0$, the negativity (N) of the state is zero, but the CCNR criterion exists. That means for $D=0$, the state is bound entangled, and no free entanglement exists in the state. This result signifies that for the specific value of the interaction strength $D=0$, the operators $O_{1}$ and $O_{2}(t)$ commutes with each other, and no QI scrambling takes place in the state. As the value of the interaction strength $D$ is introduced in the state, the negativity (N) of the state increases with time and attains a maximum value of around $0.1$. This phenomenon shows that the free entanglement develops in the state with the introduction of the interaction strength $D$. For the further increment of the interaction strength $D$, oscillatory attitude raises in the system with growing frequency. As a result, negativity (N) attains its maximum value more quickly but remains approximately stable with a minor trace of disturbance, which can be verified from figure \ref{f3}.

\subsection*{Case 2: Effect on Jurkowski \textit{et al.} state}
\label{sec:6}
In this case, a parameterized bound entangled state, proposed by Jurkowski \textit{et al.}, is adopted and talk over in the equation \ref{Js}. The selected state is shaped with three parameters and appears with different parameter conditions, which are shown in figure \ref{f1}. In the present study, the parameter condition of the state is set as $\epsilon_{1}=1;\epsilon_{2}=\epsilon_{3}=\epsilon$ and passed through the evolution process. For the parameter value $\epsilon = 4$ and different values of the interaction strength $D$, the results are shown in figure \ref{f4}, and for the interaction strength $D=0.6$ and multiple values of the parameter $\epsilon$, the results are shown in figure \ref{f5}.

Figure \ref{f4} shows that for a particular value of the parameter $\epsilon$ and the interaction strength $D=0$, the CCNR criterion exists, but the negativity (N) of the state is zero. This result also indicates that the state is bound entangled without any free entanglement for the considered value of the interaction strength $D$. As the interaction strength $D$ is introduced, a smooth free entanglement is raised in the state with the increment of negativity (N) with time and attains the ceiling value around $0.1$ as in the previous case. Further advancement of the interaction strength $D$ enhances the frequency of the oscillatory behavior of the system for which the maximum value of negativity reaches more quickly and a small amount of disturbance is generated in the negativity (N).

In figure \ref{f5}, the interaction strength $D$ is fixed at $D=0.6$ and the parameter $\epsilon$ varies. For the parameter $\epsilon=0.5$, it is found that negativity (N) and CCNR criterion both attend in the state, which implies that the bound entanglement and the free entanglement both exist in the state for the selected parameter value. From the discussion of the previous section, it is found that the state is separable for the parameter value $\epsilon = 1$, which is shown in figure \ref{f12}. On the contrary, figure \ref{f5} depicted that negativity (N) and CCNR criterion are incremented in the state with the forward movement of time for the parameter $\epsilon = 1$ and the interaction strength $D=0.6$. This result concludes that the bound entanglement and the free entanglement both are developed in the separable portion of the state, and the state becomes entirely entangled. The maximum value of negativity (N) and CCNR criterion is amplified very slowly with the further increment of the value $\epsilon$, which can be observed in figure \ref{f5}.

\begin{figure*}
\centering
\includegraphics[scale=0.98]{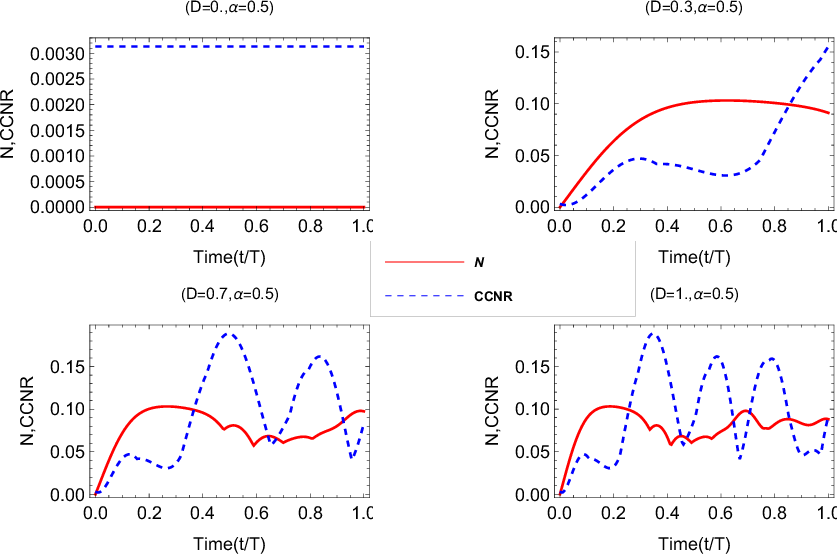} 
\caption{Plot of Negativity(N) and CCNR vs. Time$(t/T)$ for $\alpha =0.5$ and different values of $D$}\label{f6}
\end{figure*}

\subsection*{Case 3: Effect on Horodecki \textit{et al.} states}
\label{sec:7}
In the current case, two bound entangled states are considered. Both the states are proposed by Horodecki \textit{et al.} The considered states are discussed in the previous section and depicted their behavior with the parameter $\alpha$ in figure \ref{f2}. The effect of QI scrambling on both states is described below in the given successive subsections

\subsubsection*{Effect on State 1:}\label{sec:8}
\begin{figure*}
\centering
\includegraphics[scale=1]{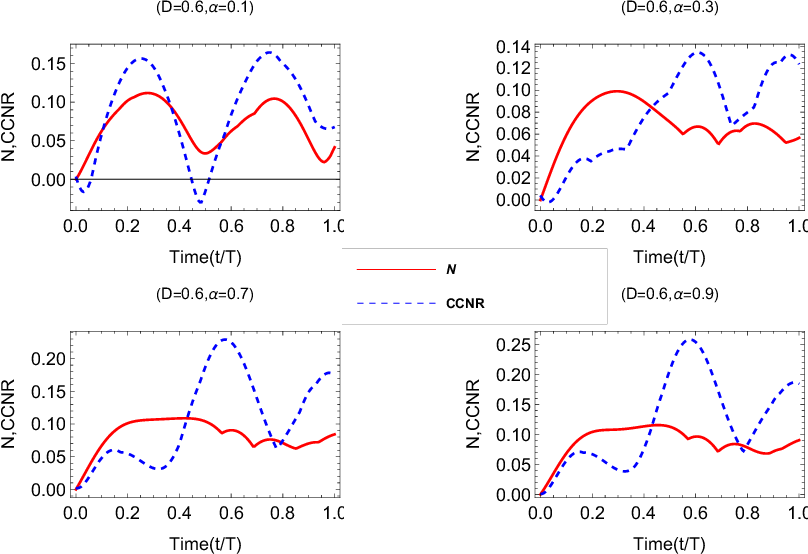} 
\caption{Plot of Negativity(N) and CCNR vs. Time$(t/T)$ for $D=0.6$ and different values of $\alpha$}\label{f7}
\end{figure*}

The first bound entangled state \textbf{[State 1]} proposed by Horodecki \textit{et al.} is described in equation \ref{hs1} and selected here to discuss the effect of the QI scrambling on it. After passing through the evaluation process, the outcomes are depicted in Figs. \ref{f6}, \ref{f7}. For the parameter value $\alpha = 0.5$ and different values of the interaction strength $D$, the results are displayed in figure \ref{f6} and for the interaction strength $D=0.6$ and different values of the parameter $\alpha$, the results are exhibited in figure \ref{f7}.

Figure \ref{f6} depicts the same behavior as the previous case for a particular value of the parameter $\alpha$ and the interaction strength $D=0$ and manifested the same conclusion that in the absence of the interaction strength $D$, the state is bound entangled without any free entanglement. With the introduction of the interaction strength $D$, negativity (N) is raised with time and achieves the maximum value around $0.1$. Further increment of the interaction strength $D$ is responsible for increasing the frequency of the oscillatory nature of the system, which made the negativity (N) of the system more unstable.

Figure \ref{f7} has displayed the outlook of the state for a particular value of the interaction strength $D=0.6$ and different values of the parameter $\alpha$. The figure indicates that the initial values of the parameter $\alpha$ negativity (N) and CCNR criterion exhibit the oscillatory behavior, which is sinusoidal in nature. As the grade of the parameter $\alpha$ increases, the oscillatory pattern of negativity (N) moves toward stability with an amount of distortion, which can be shown in figure \ref{f7}.

\subsubsection*{Effect on State 2:}\label{sec:9}
\begin{figure*}
\centering
\includegraphics[scale=1]{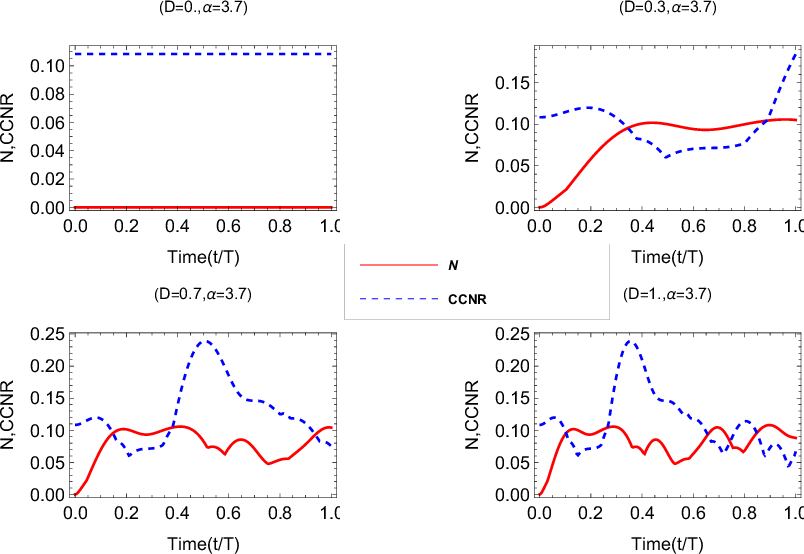} 
\caption{Plot of Negativity(N) and CCNR vs. Time$(t/T)$ for $\alpha =3.7$ and different values of $D$}\label{f8}
\end{figure*}

\begin{figure}
\centering
\includegraphics[scale=1]{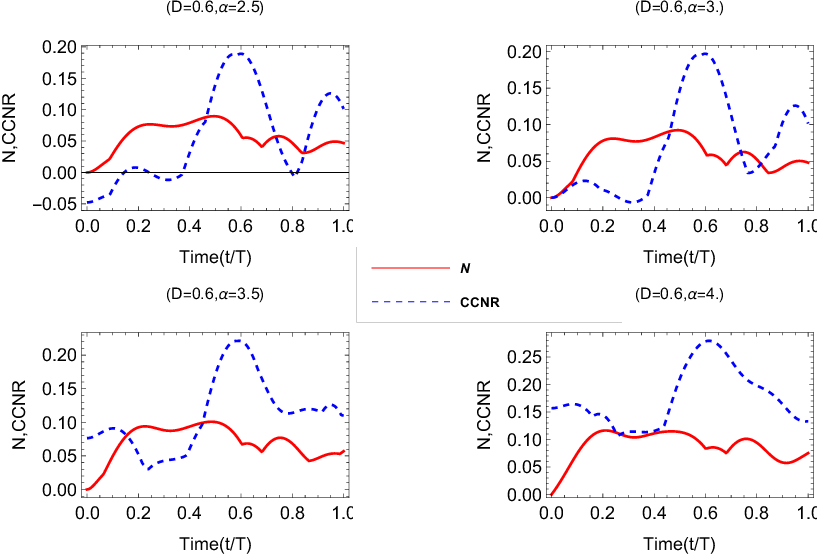} 
\caption{Plot of Negativity(N) and CCNR vs. Time$(t/T)$ for $D=0.6$ and different values of $\alpha$}\label{f9}
\end{figure}

The effect of the QI scrambling on the second bound entangled state \textbf{[State 2]} proposed by Horodecki \textit{et al.}, described in equation \ref{hs2}, is discussed here. After passing through the evaluation process, the outcomes are exhibited in Figs. \ref{f8}, \ref{f9}. For the parameter value $\alpha = 3.7$ and discrete values of the interaction strength $D$, the outcomes are shown in figure \ref{f8} and for the interaction strength $D=0.6$ and different values of the parameter $\alpha$, the outcomes are shown in figure \ref{f9}.

For a particular value of the parameter $\alpha$ and the interaction strength $D=0$, figure \ref{f8} depicted the same behavior as the previous cases and proclaimed that in the absence of the interaction strength $D$, only the bound entanglement exists in the state without any free entanglement. With the introduction of the interaction strength $D$, negativity (N) increases with time and achieves the maximum value of around $0.1$. With the further increment of the interaction strength $D$, the frequency of the oscillatory behavior of the system increases. As a result, negativity (N) gains its maximum value more quickly, and the non-sinusoidal oscillatory behavior in the system develops distortion in the negativity (N) that can be noticed in figure \ref{f8}.

Figure \ref{f9} displays the behavior of the state for the interaction strength $D=0.6$ and different values of the parameter $\alpha$. The figure shows that for the parameter value $\alpha = 2.5$, the negativity (N) increases in the state with the advancement of time and attains the maximum value around $0.1$, the same as the previous cases. This outcome implies that free entanglement developed in the state with time. On the other hand, according to the previous discussion, it has been found that for the value of the parameter $\alpha = 2.5$, the state is separable and has already been shown in figure \ref{f22}. From this discussion, it can be accomplished that the free entanglement develops in the separable part of the particular state and makes the whole state entangled. Further increment of the parameter $\alpha$ carries the same attitude of the negativity (N) for the remaining range and doesn't affect the state significantly, which can be seen in figure \ref{f9}.

\section{Conclusion}\label{sec:10}
In this article, the effect of QI scrambling is cultivated on the bound entangled states. The study is conducted on the four different bound entangled states proposed by Bennett \textit{et al.}, Jurkowski \textit{et al.}, and Horodecki \textit{et al.} During the study, the swap operator is selected as the evolution operator and the operator evolved under DM interaction. The effect of QI scrambling on each bound entangled state is described in the respective cases with a complete detailed analysis. Analyzing all the cases, it has been found that although QI scrambling minimizes the quantum information by reducing the free entanglement of the systems, using the considered operator and interaction, QI scrambling can activate the bound entangled states by introducing a certain amount of approximately stable free entanglement. It is also detected that, due to QI scrambling, both free entanglement and bound entanglement are developed in the separable portion of the selected bound entangled states and make the states entirely entangled for the whole parameter range. Further, the study can be continued for different operators and interactions to understand the behavior of multiple bound entangled states under QI scrambling.

\end{document}